# Study on the atmospheric pressure homogeneous discharge in air assisted with a floating carbon fibre microelectrode

Shuai Zhao[1(a)], Yong Hao[1], Tao Qiao[1], and Weisheng Cui[2(b)]

[1] *Aerospace Information Research Institute, Chinese Academy of Sciences - 100094, Beijing, China*
[2] *College of Physics and Optoelectronic Engineering, Shenzhen University, - 518060, Shenzhen, China*

**Abstract** – Based on the consideration of increasing the number of initial electrons and providing an appropriate distribution of electric field strength for discharge space, a method of adopting the wire-cylindrical type electrode structure with a floating carbon fibre electrode to achieve atmospheric pressure homogeneous discharge (APHD) in air is proposed in this paper. Studies of the electrode characteristics show that this structure can make full use of the microdischarge process of the carbon fibre microelectrode and the good discharge effect of the helical electrode to provide plenty of initial electrons for the discharge space. Besides, the non-uniform electric field distribution with gradual change is conducive to the slow growth of electron avalanches in this structure. Thus, the initial discharge voltage can be reduced and the formation of filamentary discharge channels can be inhibited, which provides theoretical possibilities for the homogeneous discharge in atmospheric air. Experiments show that a three-dimensional uniform discharge phenomenon can be realized under a 6.0 kV applied voltage in a PFA tube of 6 mm inner diameter, displaying good uniformity and large scale.

**Introduction.** – The low-temperature plasma produced in atmospheric pressure, especially in air, can be applied to various industrial production processes and a range of advanced applications, such as environmental protection, material science and health care, which is very promising[1-5]. To avoid the low efficiency of corona discharge and the damage of arc discharge to samples, homogeneous discharge plasma with moderate power density (such as glow or glow-like) in atmospheric air should be the best choice. But it is difficult to prevent the discharge from a filamentary mode.

In order to avoid the filamentary discharge (streamer formation) at high pressure, according to the theory of gas discharge, it is necessary to control the development of electron avalanches to prevent them from growing rapidly, so inhibiting the production of secondary electrons using a dielectric barrier discharge (DBD), it is relatively easy to generate atmospheric pressure homogeneous discharge (APHD) plasma in some special conditions, such as rare gases, radiofrequency, special barrier materials and so on [6-9].

Besides, researchers believed that one way to obtain the slow growth electron avalanches was to provide more electrons to the discharge space in an electric field of low intensity, so it was expected to form a uniform discharge in atmospheric air via reducing the impact ionization coefficient. Many researchers carried out relevant experiments. Wang *et al*. studied the influence of air gap size on the discharge process through experiments and judged that atmospheric pressure glow discharge (APGD) could be realized in the air gap with a length of no more than 2 mm. They thought that if the electric field strength of discharge space was not reduced, for the air gap with a length greater than 5 mm, the discharge had to be in the streamer form [10]. Xian and Zhou *et al* expanded the discharge air gap to 7 mm [11]. Although they was confined to generating a two-dimensional uniform air plasma sheet, their methods of controlling the development rate of electron avalanche by the high density of seed electrons and the reduced number density of molecules in air had guiding significance for expanding the three-dimensional size of the discharge gap. Liu *et al* explored the generation of APGD plasma in air based on the electric field design and successfully applied the results to the surface modification of polymers [12, 13].

In fact, the essence of almost all homogeneous discharge methods in atmospheric pressure air is to inhibit filamentary discharge by using the DBD mode to limit the development of discharge current and by providing sufficient and evenly distributed initial electrons for the discharge space to reduce the initial discharge voltage [14]. In the process of discharge, the number of initial electrons and the distribution of electric field strength are the two decisive factors affecting the uniformity of discharge. Based on the above ideas, in the previous research, we have realized the diffuse glow discharge in atmospheric air by adopting the carbon fibre helical-contact electrode structure [15]. In this paper, with the help of the floating carbon fibre microelectrode, we can make full use of advantages of the helical-contact electrode structure without additional power supply and restrain the development of electron avalanches, so as to achieve a three-dimensional uniform discharge phenomenon in large-scale space.

[a] E-mail: zhaoshuai@aoe.ac.cn (corresponding author)
[b] E-mail: wshcui@szu.edu.cn
.

**Experimental setup.** – The experimental system includes three components: the power supply, the electrode structure and the measuring system, as shown in Fig. 1.

For the whole electrode structure, it consists of two parts: a carbon fibre helical-contact electrode and a PFA tube. By winding the insulated electrode (a metal wire covered with a PTFE dielectric) with the carbon fibre electrode at a certain angle and bringing them in close contact with each other, a helical-contact electrode structure is formed. The outer diameter of the metal wire and the thickness of PTFE dielectric are both 0.2 mm, and the length of them is at least 3 cm. Carbon fibre bundle (Tenax HTA40 1K) for the helical electrode is produced by Toho Co., with a resistivity of $1.6 \times 10^{-3}$ Ω·cm. Tufted carbon fibre helically twines around the dielectric with a pitch of 5 mm. The carbon fibre helical-contact electrode is at the centre of the PFA tube along the axial direction. The PFA tube, whose inner diameter and outer diameter are 6 mm and 8 mm respectively (i.e., the wall thickness is 1 mm), is wrapped with an aluminium foil.

The power supply is a high-frequency, high-voltage sine wave generator with a voltage range of 0-10 kV and a frequency of 20 kHz. The metal part of helical-contact electrode structure is connected with the high-voltage terminal of the output, while the carbon fibre part is not electrically connected, and the aluminium foil on the PFA tube is grounded.

The discharge voltage is measured with a high-voltage probe (Tektronix P6015A), and the discharge current is obtained by measuring the voltage across a resistance $R_m$ in series with the electrode. The waveforms of discharge voltage and current are recorded using a Tektronix digital oscilloscope (TDS1012B-SC).

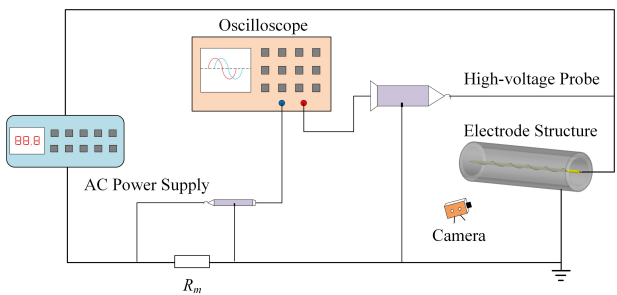

Fig. 1: (Colour on-line) Experimental system for discharge.

**Results and discussions.** –

*Discharge characteristics of the carbon fibre helical-contact electrode.* The carbon fibre helical-contact electrode structure is sketched in Fig. 2(a). To conduct a better simulation study, the carbon fibre bundle is taken as an integrated whole and the outer diameter of it is set to 0.2 mm. With the above setup configured in the electromagnetic simulation software Ansoft Maxwell, when a 1.8 kV voltage difference is applied between the carbon fibre electrode and the metal electrode, we can obtain the corresponding cross section of the electric field distribution as shown in Fig. 2(b). The whole electrode structure is wrapped by a strong electric field region distributed as a wavy profile. The wrapping shape caused by the electric field distribution is favour of the charged particles to drift and diffuse in the three-dimensional space, which creates the conditions for the generation of APGD plasma in air.

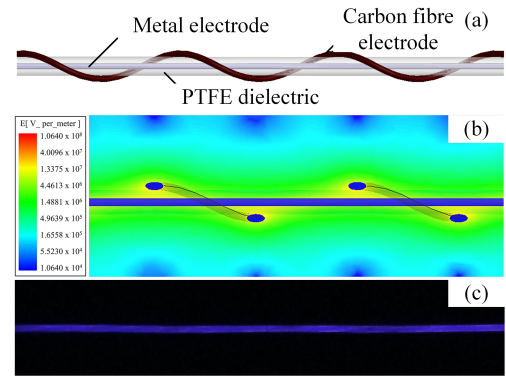

Fig. 2: (Colour on-line) The theoretical features and discharge characteristics of the carbon fibre helical-contact electrode. (a) the structure of the carbon fibre helical-contact electrode; (b) the distribution characteristics of the electric field strength; (c) the discharge phenomenon under a voltage of 1.8 kV (exposure time: 1/12 s).

The actual monofilament of the carbon fibre is only 7 μm, so the initial discharge space is limited to a microscale under this microelectrode, thus producing a microdischarge (at least one dimension of the discharge space is below a millimeter scale). For sufficiently small gaps, especially for gap distances below 10 μm, the field emission is dominant over Townsend discharge [16-18]. The combination of field emission and collisional effects reduces the breakdown voltage for decreasing gap distance [17].

The threshold electric field required for field emission is a function of the cathode material and surface properties [16]

$$D_{FN} = (6.85 \times 10^7) \frac{\phi^{3/2}}{\beta} \quad [\text{V/cm}], \quad (1)$$

where $D_{FN}$ is the threshold electric field, $\Phi$ the work function of the cathode material, and $\beta$ the geometric enhancement factor dependent on the geometry of the electrode and the electrode gap distance. The threshold electric field of carbon fibre is low, generally in the range of $1\times10^6$ - $3\times10^6$ V/m [19]. In the small gap near the contact points of the carbon fibre helical-contact electrode, the field strength obviously reaches this value. The work function of carbon fibre is 4.7 eV, which is approximately the same with copper, so $\beta$ of the carbon fibre can be estimated via using Eq. (1). It is about in the range of $2.33\times 10^4$ - $6.98\times10^4$, which is so big. Therefore, the large enhancement factor enables the field emission to be achieved at much lower field strength and confirms that the shape of the fibre electrode (sufficiently small curvature radius) and the small electrode gap have great advantages. A great quantity of initial electrons are filled in the discharge space owing to the strong field emission under the microdischarge, effectively reducing the breakdown voltage and easily realizing homogeneous discharge under relatively low average field strength, so it can be predicted that the microdischarge will play a triggering and guiding role in the generation of the well-distributed plasma.

In the discharge experiment, it was found that the initial discharge voltage of carbon fibre helical-contact electrode was only 960 V. Fig. 2(c) illustrated the discharge phenomenon when the applied voltage was up to 1.8 kV. A three-dimensional and homogeneous glow discharge emerged from the surface of the entire electrode. The glow diffused along the radial direction of the wire electrode and completely wrapped it with plasma.

The good discharge phenomenon of the carbon fibre helical-contact electrode confirms that the microdischarge caused by the carbon fibre filaments greatly reduces the threshold discharge voltage; so that the electrode gap can have many initial electrons at a low average voltage, triggering the whole discharge process of the helical electrode to further produce a large number of electrons. Meanwhile, the uniformity of the discharge can be guaranteed by the special electric field distribution of the carbon fibre helical-contact electrode.

*Analysis of the wire-cylindrical electrode structure with floating carbon fibre.* As described in *Experiment setup*, the whole wire-cylindrical electrode structure is sketched in Fig. 3. When the metal electrode in the carbon fibre helical-contact electrode structure and the aluminium foil electrode on the outer surface of the tube are connected with the high-voltage terminal and the ground terminal respectively, the carbon fibre is in the floating state.

Considering that the time for the conductor to reach electrostatic equilibrium in electrostatic field is very short (about $10^{-14}$ s), far shorter than the reversal time of general sine wave power supply (for example, the reversal time of the 20 kHz power supply is $10^{-5}$ s), the conductor floating in the sinusoidal electric field is similar to that in the electrostatic field. It can get an induced potential, and the magnitude of the induced potential is related to the field strength of the conductor's position. When the voltage $U$ is applied to the pair of electrodes, the electric field $E_{tube}$ is generated in the tube space. Although the carbon fibre is in floating state, it can get the induced potential under the action of $E_{tube}$ and form the potential difference $U'$ with the high-voltage electrode. The higher the applied voltage $U$ is, the greater the potential difference $U'$ becomes.

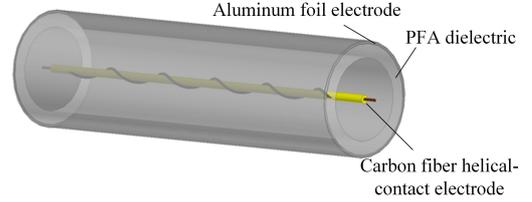

Fig. 3: (Colour on-line) Schematic diagram of the wire-cylindrical electrode with floating carbon fibre

Based on the principles of induced potential, we can calculate the induced potential of carbon fibre electrode conveniently by Ansoft Maxwell. The calculation results of potential difference $U'$ under different applied voltage $U$ are shown in Table 1.

Table 1: Relationships between the applied voltage and the induced voltage difference.

| Applied voltage $U$ (kV) | 2 | 3 | 4 | 5 | 6 | 7 |
|---|---|---|---|---|---|---|
| Induced voltage difference $U'$ (kV) | 0.50 | 0.75 | 1.00 | 1.25 | 1.50 | 1.75 |

It can be seen from the above sections and relevant references[20-23] that, due to the very small discharge gap, with the help of microdischarge and field emission, the carbon fibre helical-contact electrode structure is easy to preferentially form the discharge process under a certain electric field strength. The predischarge effect is optimized because of its low initial discharge voltage and good diffusion. Therefore, there are two main sources of initial electrons in the tube space: a) the electrons released during the microdischarge process and drifting out of the small gap under the electric field; b) the electrons generated in the glow discharge process of the carbon fibre helical-contact electrode. These initial electrons produced near the central axis of the PFA tube, as the seed electrons, will cause a new collision ionization process under the action of $E_{tube}$, which makes it possible for the uniform discharge in the whole tube space.

The increase of $U$ makes it very easy to cause filamentous discharge, so it is not necessary to make the value of $U'$ reach 1.8 kV. It's a very definite possibility to achieve homogeneous discharge in the tube space only by ensuring the number and uniformity of the initial electrons provided by the carbon fibre helical-contact electrode.

The distribution of $E_{tube}$ also has a great influence on the discharge state. When the applied voltage is 6.0 kV, we can obtain the electric field intensity distribution of the whole tube space, as shown in the radial section in Fig. 4 (a). The electric field vector radiates uniformly from the central axial of the tube to the outer wall. It is easy to imagine that, compared with the wire-plate electrode structure with the same floating carbon fibre and equal gap distance, the wire-cylindrical electrode structure, benefitting from the radial symmetry of the electric field distribution, weakens the field strength near the helical electrode and makes it more uniform. It also obviously has more dimensions in uniformity than the wire-plate electrode structure. These characteristics make it more conducive for the wire-cylindrical electrode with floating carbon fibre to form homogeneous discharge in air.

The radial electric field intensity distribution curve can be obtained by drawing a line from A to B along the inner diameter of the tube, as shown in Fig. 4 (b). It can be seen that the maximum electric field intensity in the tube space is only $6.5 \times 10^6$ V/m, and the field strength along the radial direction (such as the path from c to b) decreases gradually from the central axis of the tube. Compared with the wire-cylindrical electrode structure without floating carbon fibre, whose electric field intensity distribution and radial distribution curve (from A' to B') are shown respectively in Fig. 4(c) and 4(d), the wire-cylindrical electrode structure with carbon fibre helical electrode has a greater maximum electric field strength near the central axis, which indicates more advantages and possibilities in field emission of the carbon fibre. However, the gradient trends of reduced electric field intensity in the radial tube space are the same.

In order to verify the influence of the inhomogeneous electric field intensity on the impact ionization process, we can introduce a rough numerical calculation.

At the atmospheric pressure and room temperature, the electron collisions are frequent due to their short mean free path (it is about 0.377 μm calculated by the volume fraction of nitrogen and oxygen [15]) , so the key to restrain the development speed of avalanches is to dwindle the rate coefficient $\alpha$ for electron impact ionization. If we only consider impact ionization (the $\alpha$ process), the number of electrons follows an exponential rise

$$n = n_0 e^{\alpha d}, \qquad (2)$$

where $n_0$ is the initial number of electrons, $d$ is the development distance of the electron avalanche. In air, $\alpha$ can be calculated as [10]

$$\alpha = 6460 e^{-\frac{1.9 \times 10^7}{E}}, \qquad (3)$$

where $E$ is the magnitude of the applied electric field in range of $1.5 \times 10^6$ - $1.1 \times 10^7$ V/m, so we take the electron avalanche development in the tube space within this range of field strength as an example, i.e., the path c-d in Fig. 4(b). Taking the central axis of the tube as the coordinate origin , Fig. 5(a) shows the profiles of the electric field strength along path c-d and a nonlinear fitting result of it by the data analysis software Origin. The electric field $E$ is a function of path length $x$ represented as

$$E(x) = 1.27 \times 10^7 e^{-\frac{x}{0.315}} + 1.45 \times 10^6 \quad [\text{V/cm}]. \quad (4)$$

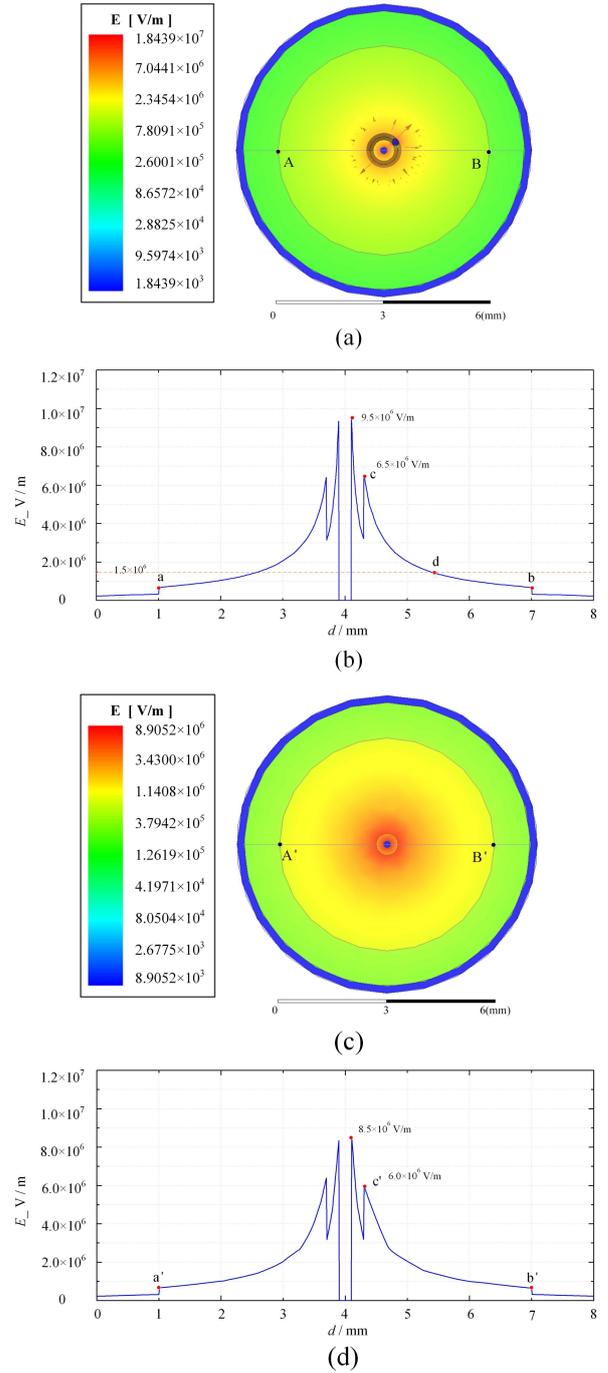

Fig. 4: (Colour on-line) Distribution characteristics of the wire-cylindrical electrode structure under the 6.0 kV applied voltage. (a) the wire-cylindrical electrode structure with floating carbon fibre; (b) the electric field intensity distribution along the line from point A to point B; (c) the wire-cylindrical electrode structure without floating carbon fibre; (d) the electric field intensity distribution along the line from point A' to point B'.

Assuming electron collisions occur along each mean free path, then from an iterative calculation using Matlab based on Eqs. (2), (3), and (4), we can obtain the development number $n$ of one initial electron in the process of impact ionization and the rate coefficient $\alpha$ along the path c-d, as shown in Fig. 5(b).

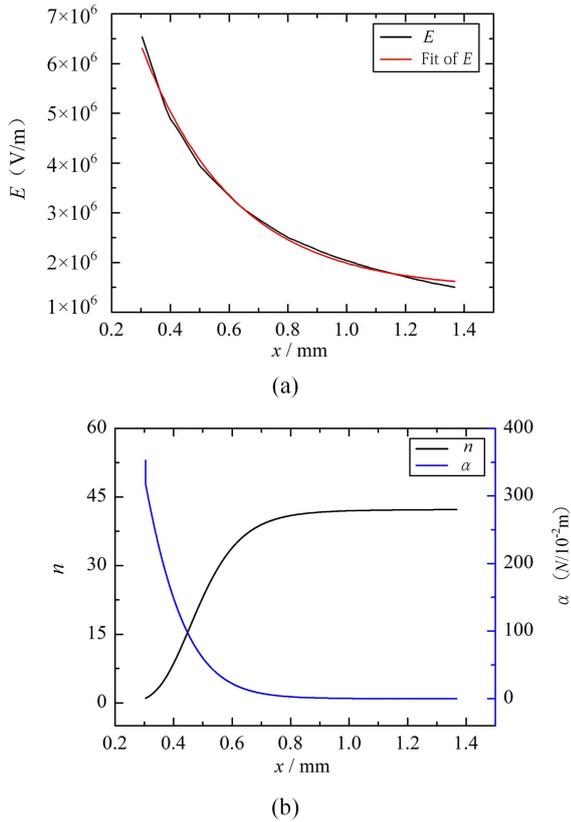

Fig. 5: (Colour on-line) Field strength distribution and development process of an electron avalanche along the path c-d. (a) the distribution of field strength along the path c-d and the result after nonlinear fitting; (b) the change of electron number and $\alpha$ when an initial develops during impact ionization along the path c-d.

It can be seen that the electron avalanches developed from the central axis area grow slowly and the number of electrons in avalanche heads increases to 42 times after a path of about 1mm. If choosing an average intensity of electric field on the path c-d ($\bar{E} = 4.02 \times 10^6$ V/m) and assuming that the impact ionization process occurs in this constant field strength and develops the same path length, we can demonstrate that 440 times electrons are produced at last, which is far greater than the number along the inhomogeneous field path c-d. This indicates that the gradient field distribution effectively suppresses the electron growth in the impact ionization process. When the carbon fibre helical electrode is used as the cathode (the applied voltage is in the negative half cycle), the electron avalanches will begin to develop from the area near the central axis. At the early stage of electron avalanches, the stronger $E$ in the tube is helpful to give rise impact ionization. When the electron avalanches grow to a certain distance, the weaker $E$ can restrain the size of electron avalanche heads. Such electric field distribution is conducive to limiting the development speed of the electron avalanches generated from the central axis of the tube, so as to form slowly growing electron avalanches and avoid the formation of filamentary discharge. When the aluminum foil electrode is used as the cathode (the applied voltage is in the positive half cycle), taken as the main source of the initial electrons, the charged particles adsorbed on the inner wall of PFA tube released into the discharge space under the action of electric field, forming the reverse electron avalanches and maintaining the discharge process in the air gap. However, due to the small $E$ near the inner wall of the tube, it is not easy for the initial electrons to obtain enough energy to cause impact ionization, so that the development paths of electron avalanches in this half cycle will not sustain too long, thus also being unable to form violent discharge channels.

When the $E$ is less than $1.5 \times 10^6$ V/m, Eq. (3) is no longer applicable. With the decrease of $E$, $\alpha$ is greatly reduced and the number of electrons in avalanche heads is difficult to continue increasing, thus reaching a "saturation state". For the low field strength area, it mainly relies on the diffusion and drift of electrons to achieve the discharge throughout the tube space.

Each electron avalanche is funnel-shaped during its development along the radial direction of the tube, that is to say, it is narrow near the cathode but wide near the anode. The initial electrons produced by the discharge of carbon fibre helical-contact electrode structure are evenly distributed along the axial direction of the tube, so that multiple overlapping electron avalanches can appear near the cathode, thus forming uniform discharge channels and preventing their contraction. The uniform distribution of discharge plasma in the tube space further weakens the spatial electric field and avoids the formation of filamentous discharge.

To sum up, by virtue of the discharge advantages of the carbon fibre helical-contact electrode structure and the characteristics of radially inhomogeneous electric field distribution in the tube space, a large number of electrons can be obtained in the discharge space under a lower average electric field intensity, and the slow growth of electron avalanches can be realized, which creates favorable theoretical conditions for the formation of homogeneous discharge in the whole tube space.

*Discharge phenomenon and characterises.* Based on the experimental setup in Fig. 1, we used NIKON COOLPIX P100 camera to capture the discharge phenomenon of the wire-cylindrical electrode with floating carbon fibre.

When the applied voltage was 3.8 kV, some luminous points emerged in the tube, which mainly located near the surface of the helical electrode, and a certain width of glow appeared along the longitudinal direction of the helical electrode. With the increase of the voltage, the width of the glow area expanded. When the voltage was up to 6.0 kV, the induced voltage difference was about 1.50 kV, which had not reached the stable discharge voltage of the helical electrode itself, the stable discharge region in the tube space was already full of diffuse and three-dimensional glow or glow-like phenomena, as shown in Fig. 6.

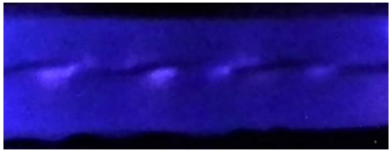

Fig. 6: The experimental phenomenon with an applied voltage of 6.0 kV (exposure time: 1/3 s).

The discharge current and applied voltage waveforms are shown in Fig. 7. There are preferential discharge points in the non-uniform spatial electric field formed by this electrode structure, thus representing a non-synchronous discharge in time and space, leading to a combination of multiple current pulses with many peaks. The discharge consists of many short (<10 ns) microdischarges, which is indicated also by the peaks seen on the discharge current waveforms [24]. Fig. 7 shows that the values of discharge currents are in the magnitude of dozens of mA , demonstrating that there does not appear lots of high concentration of plasma channels in the discharge space, and the dispersive uniform discharge state dominates in this process. However, the stable discharge current amplitude of the wire-cylindrical electrode with floating carbon fibre is larger than that of only carbon fibre helical-contact electrode (as a matter of experience, it is about 0-40 mA), which is supposed to be caused by the longer development paths of electron avalanches.

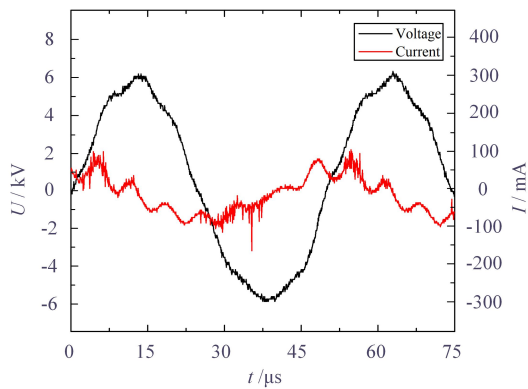

Fig.7: Waveforms of applied voltage and discharge current.

As we all know, in the electrode structure of parallel-plate or wire-plate with an air gap of no less than 3 mm, it is more likely to form filamentary discharge and corona discharge. However, in this study, a more uniform glow or glow-like discharge phenomenon is obviously observed in atmospheric air, and the steady gap voltage is about 4.67 kV, which is lower than the parallel-plate electrode and wire-plate electrode under the same conditions. It is also much lower than the static breakdown voltage (11.2 kV) in 3 mm air gap calculated from Paschen curve [6, 25]. This fully proves that the predischarge of carbon fibre helical-contact electrode and the electric field intensity with gradual change have great influence on the discharge process. On the one hand, specifically speaking, due to its characteristics (including lower discharge voltage, good diffusion and discharge uniformity), the carbon fibre helical-contact electrode initiates a predischarge at a lower voltage (induced voltage difference in this paper) and generates plasma containing a large number of electrons as seed electrons, which provides favourable conditions and plays a good trigger role for the discharge in the whole tube space. The 1.50 kV induced voltage difference can facilitate a good discharge phenomenon without reaching the stable discharge voltage of the carbon fibre helical-contact electrode itself, which just confirms this point of view. On the other hand, the radial electric field strength from the axis of the PFA tube to the inner wall of the tube presents a trend of gradual reduction, which restrains the development of electron avalanches and avoids the formation of constricted discharge channels or intense arc discharge, thus maintaining the state of uniform discharge. In the experiment of this paper, the interaction of above two points directly makes the discharge space obtain more electrons under a lower electric field intensity and develop electron avalanche process with slow growth, thus achieving a well-distributed discharge phenomenon in atmospheric pressure air.

Intuitively speaking, in atmospheric air, the larger the gap between electrodes, the more difficult it is to form uniform discharge. The atmospheric air discharge in this paper is the result of the combined action of the initial electrons and the spatial electric field intensity distribution. On the premise that the initial electron number is constant, i.e., the induced differential pressure $U'$ is constant and the predischarge capacity of the carbon fibre electrode is consistent with the foregoing experiment, the value of external voltage $U$ shall be increased if the inner diameter of the PFA tube is enlarged. Although the electric field intensity gradually decreases from the central axis to the inner wall of the tube, with the increase of the inner diameter and the applied voltage, the trend of electric field changes little, while the average electric field intensity obviously increases. If the impact ionization goes through a longer path under higher average electric field intensity, it is impossible to avoid the occurrence of violent electron avalanches and the formation of filamentary discharge. So we need to match the applied voltage, the inner diameter of the PFA tube and the discharge state of the helical electrode, and comprehensively consider variable conditions to get an applicable size of the electrode structure in this paper. We will focus on the parameter matching and more related mechanism issues in the next work.

**Conclusions.** – In summary, we have proposed a method of using carbon fibre floating electrode to form APHD in air. By designing the wire-cylindrical electrode structure with floating carbon fibre, a three-dimensional uniform discharge phenomenon in 6-mm wide cylindrical space is realized under a single applied voltage. This is mainly attributed to the following two theoretical features:

a) With the assist of the carbon fibre microelectrode, the favourable predischarge process of the carbon fibre helical-contact electrode is easily formed, which provides enough initial electrons for the discharge in the whole tube space;

b) The non-uniform electric field distribution with gradual change along the radial direction in the tube space limits the development speed of electron avalanches and effectively suppresses the generation of filamentous discharge.

These two characteristics enable the discharge space to obtain plenty of electrons at a lower intensity of average electric field, and achieve the slow growth of electron avalanches, which promoted the formation of APHD in air.

The discharge demonstrates the advantages of good uniformity, low discharge voltage and large scale, which is very suitable for air purification and material modification. In the future, the structural parameters of the whole electrode structure will be further explored and optimized to satisfy the requirements of all kinds of applications.